\documentclass[12pt,twoside]{article}
\usepackage{fleqn,espcrc1_new}
\usepackage{graphicx,rotating,cite}

\newcommand{\jpsi} {\mbox{J\kern-0.05em /\kern-0.05em$\psi$}}
\newcommand{\psip} {\mbox{$\psi\prime$}}
\newcommand{\rpsi}{\mbox{$\psi'/\psi$}}
\newcommand{\psidy}{\mbox{$\psi$/DY}}
\newcommand{\rs}{\mbox{$\sqrt{s}$}}
\newcommand{\dy}{Drell--Yan}
\newcommand{\dd}{\mathrm{d}}

\newcommand{\et}{$E_{\mathrm{T}}$}
\newcommand{\ezdc}{$E_{\mathrm{ZDC}}$}
\newcommand{\pt}{$p_{\mathrm{T}}$}
\newcommand{\xf}{$x_{\mathrm{F}}$}

\begin{document}

\renewcommand{\textfraction}{.1}    
\renewcommand{\bottomfraction}{.85} 

\begingroup
\thispagestyle{empty}
\baselineskip=14pt
\parskip 0pt plus 5pt

\begin{center}
{\large EUROPEAN ORGANIZATION FOR NUCLEAR RESEARCH}
\end{center}

\begin{flushright}
CERN--PPE\,/\,96--158\\
LIP\,/\,96--03\\
1 November 1996
\end{flushright}
\bigskip

\begin{center}
\Large
\textbf{Hard probes in nucleus-nucleus collisions}
\end{center}

\bigskip\bigskip
\begin{center}
C.~Louren\c{c}o\\
\bigskip
\emph{CERN PPE, CH-1211 Geneva 23, Switzerland}\\
\end{center}
\bigskip\bigskip\bigskip

\begingroup
\leftskip=0.4cm
\rightskip=0.4cm
\parindent=0.pt
\begin{center}
ABSTRACT
\end{center}

  The present knowledge on hard processes in the context of heavy ion
  collisions is reviewed, with particular emphasis on \jpsi\ 
  production.  The p-A data on charmonia production from Fermilab
  experiments is shown to be in excelent agreement with the p-A data
  collected at CERN.  The simultaneous analysis of all existing p-A
  data reaches a precision which allows us to rule out some
  preconceived ideas, setting a good frame against which the data
  collected with ion beams at CERN can be compared.

\endgroup

\vfill
\begin{center}
  
  Summary talk presented at the 12th International Conference on
  Ultra-Relativistic Nucleus-Nucleus Collisions (Quark Matter '96),
  Heidelberg, Germany, May 1996.

\end{center}

\endgroup


\newpage
\pagenumbering{arabic}
\setcounter{page}{1}

\title{Hard probes in nucleus-nucleus collisions}
     
\author{Carlos Louren\c{c}o\\[3mm]
CERN PPE, CH-1211 Geneva 23, Switzerland}

\maketitle

\begin{abstract}
  The present knowledge on hard processes in the context of heavy ion
  collisions is reviewed, with particular emphasis on \jpsi\ 
  production.  The p-A data on charmonia production from Fermilab
  experiments is shown to be in excelent agreement with the p-A data
  collected at CERN.  The simultaneous analysis of all existing p-A
  data reaches a precision which allows us to rule out some
  preconceived ideas, setting a good frame against which the data
  collected with ion beams at CERN can be compared.
\end{abstract}

\section{Introduction}

The first data on \jpsi\ production in nucleus--nucleus collisions
were collected 10 years ago, at the CERN SPS, by the NA38 experiment,
with the 200~GeV per nucleon oxygen beam.

Although the experiment had been planned to search for thermal dimuons,
a paper by Matsui and Satz~\cite{Matsui_Satz} redirected 
attention by concluding that ``there appears to be no mechanism for
\jpsi\ suppression in nuclear collisions except the formation of a
deconfining plasma, and if such a plasma is produced, there seems to
be no way to avoid \jpsi\ suppression.''

In fact, already at QM~'87, the NA38 collaboration presented evidence
of a substantial decrease of the ratio $\psi$/Continuum, both from p-U to
O-U collisions and versus \et\ within O-U~\cite{Bussiere87}.  These
observations were soon confirmed with S-U data~\cite{Baglin91a}.

Although the exciting interpretation of this ``\jpsi\ suppression'' as
a signal of QGP formation was always present, some alternative
explanations of a more ``classical'' nature were soon proposed.

Several developments helped to clarify the interpretation of the data:

\begin{itemize}
  
\item Concerning the absolute cross sections, Fermilab experiment E772
  observed that \jpsi\ production is, in p-A collisions, proportional
  to A$^\alpha$, with an $\alpha$ value smaller than the one
  previously reported by NA3 (0.94$\pm$0.03~\cite{NA3_psi}) and used in
  early NA38 papers~\cite{NA38_et}.
  
\item Several measurements of $B_{\mu\mu}\sigma^\psi$ in p-A collisions were
  obtained by NA38, leading to a much more complete and accurate
  reference in the study of the ion data.
  
\item Higher statistics in the S-U data set allowed to use the high
  mass \dy\ as the reference in the \et\ dependent (relative) studies,
  replacing $\psi$/Cont by $\psi$/DY.  Furthermore, the isospin
  correction ($\sim$~10\,\%) required when comparing the proton to the
  (isoscalar) sulphur induced collisions lead to a smaller
  ``suppression'' relative to what had been reported before, using
  $\psi$/Cont. 
  
  At the same time, information on the \psip\ became available,
  revealing a ``suppression'' much beyond the nuclear absorption expected
  from the p-A data.
  
\item The absolute cross sections of direct \jpsi\ and \psip\ production were
  found to be far too high relative to the values calculated with the
  colour singlet model.  Although this was known to be the case
  already from fixed target data~\cite{Sridhar96}, the new (high \pt)
  data from CDF~\cite{CDF}, clearly identifying the fraction of \jpsi\ 
  coming from beauty and $\chi_c$ decays, triggered a major change in
  the theoretical understanding of charmonia
  production~\cite{Braaten}.  Contrary to the paradigm of the colour
  singlet model, a substantial fraction of the \jpsi\ and \psip\ cross
  section is now believed to come from binding of $c\bar{c}$ pairs
  produced in a colour octet state.
  
\end{itemize}

Taking into account all these developments, some of them only
available after QM~'95, Kharzeev and Satz~\cite{Kharzeev_Satz96} were
able to explain the phenomenological fit made earlier by Gerschel and
H\"ufner~\cite{Gerschel_Huefner}, by associating the absorption cross
section of around 6--7~mb to the $c\bar{c}g$ state crossing
the nuclear matter, rather than to a fully formed \jpsi, for which
they had previously calculated a much smaller cross
section~\cite{Kharzeev_Satz94}.

The further suppression of the \psip\ observed in S-U collisions is
explained~\cite{Kharzeev_Satz96} by the formation of a (confined)
hadronic medium in the later stages of the collision.  These
``comovers'' can easily break the loosely bound, and by then fully formed, 
\psip\ while leaving the \jpsi\ unchanged, since confined gluons are
too soft to overcome the \jpsi's ($\sim$~640~MeV) binding energy.

\bigskip

This was the status of the charmonia front just before the results
from the Pb data collected in December 1995 became available.  These
new results~\cite{Gonin} certainly open another very exciting chapter
in the \jpsi\ saga, and the people attending QM~'96 will remember the
lively discussions held inside and outside of the conference room.

Some of the proposed ``explanations'' are not accurate enough to be
seriously considered.  For instance, it is not reasonable to reproduce
the Pb-Pb data at the price of missing the p-A and S-U points.  Also,
we cannot mix the $\psi$/Cont values of S-U (presented for the last
time at QM~'91) with the $\psi$/DY values of Pb-Pb, the latter
variable certainly being much better (remember the enhancement
observed in the intermediate mass region~\cite{IMR}, which might
affect the ``continuum'').

The precision and diversity of the measurements now available
certainly deserves equally accurate theoretical descriptions.  The
field of \jpsi\ production in heavy ion collisions has reached
maturity and QM~'96 should be considered as a transition point: the
time has come to build realistic theoretical models that reproduce the
data with an accuracy at the few percent level.

In this sense, it would be very helpful to develop a microscopic model
(like RQMD, VENUS or HIJING) that would generate hard processes in a
nuclear environment, including nuclear geometry effects, modifications
of the parton distribution functions, etc.  Of course, it would have
to work also at SPS energies, not only for RHIC and LHC.

In the remaining part of this paper, the data presently available is
presented in detail, both the directly measured values and, where they
are required, those obtained after correcting for \rs\ dependences and
phase space windows.  Special emphasis is given to the p-A data,
collected at FNAL and at CERN, enough by it self to rule out some
ideas.

\section{Hard processes in p-A collisions}

\subsection{\dy\ production}

\begin{figure}[t]
\centering
\resizebox{0.85\textwidth}{!}{%
\includegraphics*{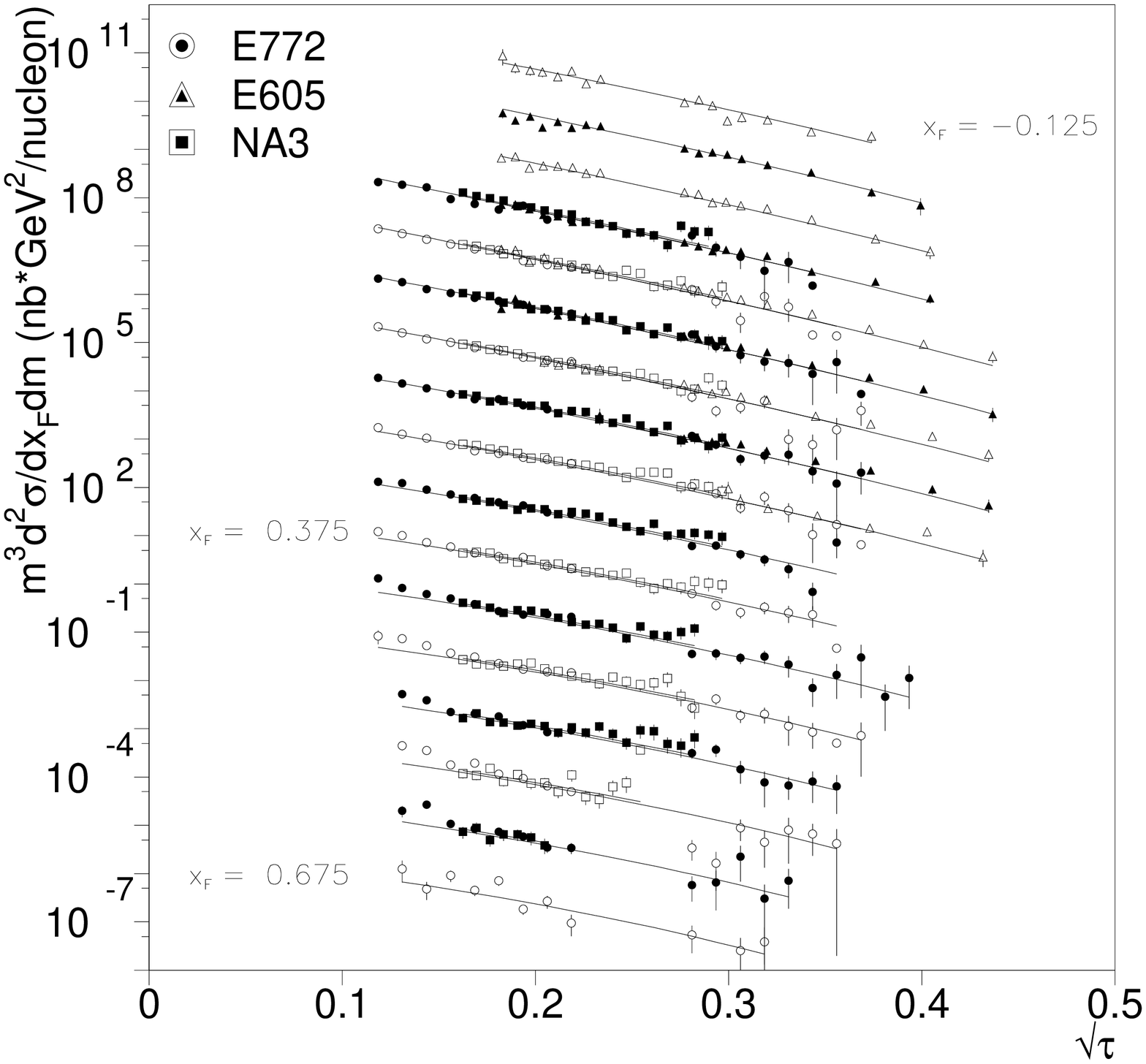}}
\caption{p-A \dy\ double differential cross section, per target
  nucleon.  See text for details.}
\label{fig:pat_dy}
\end{figure}

Figure~\ref{fig:pat_dy} presents the double differential \dy\
cross section, $m^3\,\dd^2\sigma/\dd m\,\dd x_F$, in
nb$\times$GeV$^2$/nucleon, versus the scaling variable
$\sqrt{\tau}=m/\sqrt{s}$ (using mass bins of 0.5~GeV), for different
\xf\ bins (0.05 wide).  For presentation purposes, the points above
(below) the $x_F=0.375$ data set have been successively multiplied
(divided) by factors of 10.  This figure merges \dy\ data from three
p-A dimuon experiments: E772, E605 and
NA3~\cite{E772_dy,E605_dy,NA3_dy}.  The data are available as tables
from the \textsc{hepdata} data base~\cite{hepdata}.  The beam energy
was 800~GeV in the Fermilab experiments and 400~GeV in the case of
NA3.  Different targets were used: hydrogen (E772), copper (E605) and
platinum (NA3).  The agreement between the three experiments is
excelent.  The curves on the figure are the result of a leading order
calculation using the CTEQ2M set of (NLL) quark distribution
functions~\cite{Pat}, multiplied by an appropriate K factor (around
1.5).  The K factor is considerably reduced when the calculations are
performed in NLO and becomes $\sim$~1 at NNLO~\cite{dy_th}.

The influence of the nuclear target, usually referred to as ``the
A-dependence'', is quite simple in the case of \dy\ production.  The
cross section just scales linearly with the number of target nucleons,
$\sigma^{DY}_{pA}=\sigma^{DY}_0 \times A$.  A closer look reveals that
we have to count the number of protons and neutrons separately, since
the \dy\ cross section depends on the charge of the annihilating
quarks.  In fact, the precision reached in \dy\ measurements is
such that it has revealed differences between the structure functions
of the $\bar{u}$ and $\bar{d}$ sea in the nucleons~\cite{udbar}, a
probable explanation for the violation of the Gottfried sum rule
recently observed in deep-inelastic lepton scattering~\cite{NMC}.

\begin{figure}[htb]
\centering
\resizebox{0.54\textwidth}{!}{%
\includegraphics*[bb= 38 100 556 670]{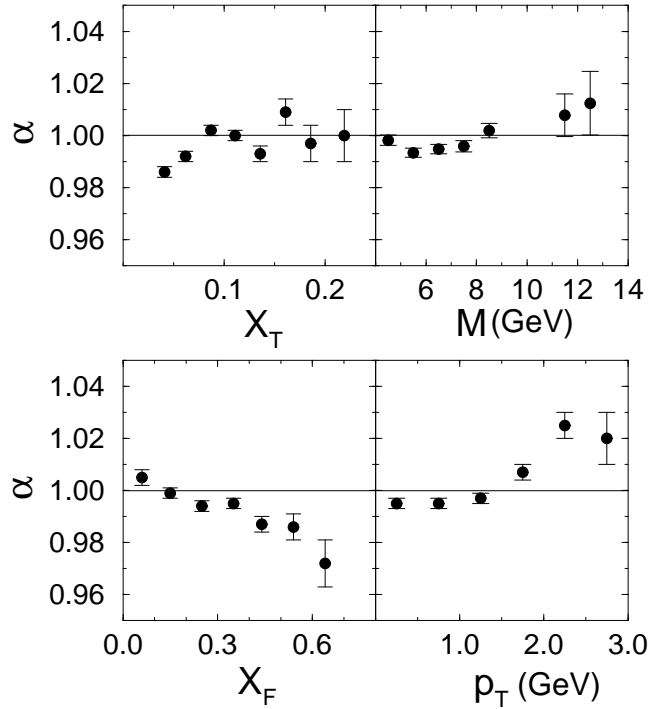}}
\caption{\dy\ A-dependence expressed as $\alpha$ versus
  $x_\mathrm{T}$, m, \xf\ and \pt.  Notice the fine scale used
  in the vertical axis.}
\label{fig:alpha_dy}
\end{figure}

The dependence of \dy\ on A, for p-A collisions, can be
expressed in terms of the $\alpha$ parameter, as done in
figure~\ref{fig:alpha_dy}, or by the ratios between the dimuon yields
per nucleon in heavy targets relative to a light target, say
deuterium, as presented in table~\ref{tab:adep_E772}.  

\begin{table}[hbt]
\caption{Ratios of DY, \jpsi\ and \psip\ cross sections between p-A and
  p-$^2$H, from E772 data.}
\label{tab:adep_E772}
\begin{tabular*}{0.85\textwidth}{@{}l@{\extracolsep{\fill}}ccc}
\hline
Target (A) & DY & \jpsi & \psip \\ \hline
C~~(12)&1.003$\pm$0.0085&0.851$\pm$0.0125&0.855$\pm$0.0292\\
Ca~(40)&0.995$\pm$0.0061&0.806$\pm$0.0087&0.750$\pm$0.0271\\
Fe~(56)&0.990$\pm$0.0065&0.756$\pm$0.0102&0.722$\pm$0.0357\\
W~(184)&0.986$\pm$0.0083&0.619$\pm$0.0133&0.623$\pm$0.0331\\ \hline
\end{tabular*}
\end{table}

\subsection{\jpsi\ production}

Table~\ref{tab:adep_E772} also includes the values measured by the
E772 experiment~\cite{Alde91} for the \jpsi\ and \psip\ resonances,
first shown at the QM~'90 conference.

\begin{figure}[htb]
\centering
\resizebox{0.5\textwidth}{!}{%
\includegraphics*[bb= 36 233 256 500]{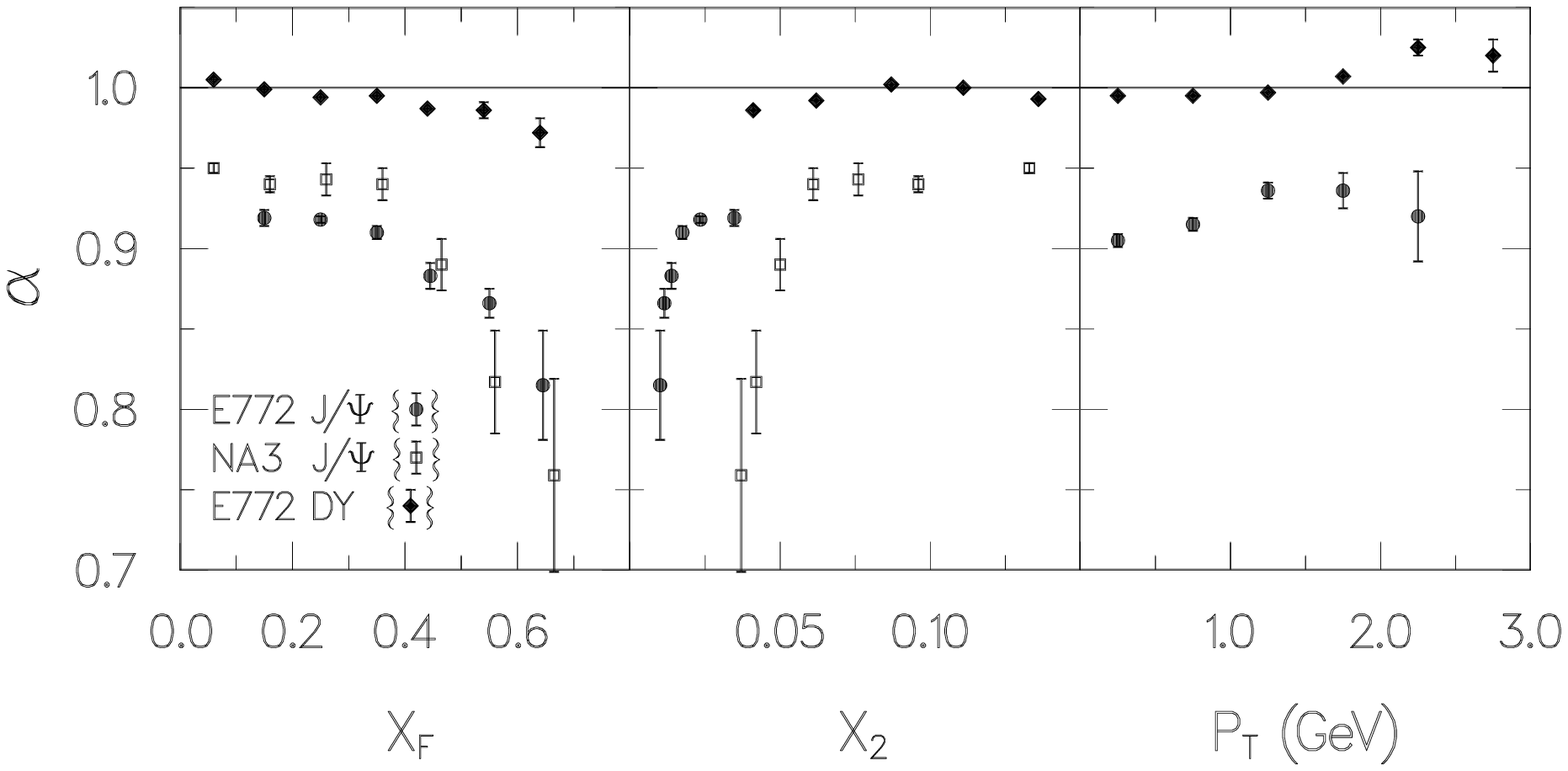}}
\caption{\xf\ dependence of $\alpha^{DY}$ and $\alpha^\psi$.}
\label{fig:alpha_xf}
\end{figure}

Before comparing the results of the CERN experiments, NA38 and NA51,
with the E772 data, in what concerns p-A collisions, we should notice
that the $\alpha$ of $\jpsi$ depends on \xf, as is shown in
figure~\ref{fig:alpha_xf}~\cite{Alde91}.  The E772 data on
$\alpha^\psi$ is also presented in table~\ref{tab:alpha_psi}.

\begin{table}[hbt]
\caption{Values of $\alpha^\psi$ versus \xf, from E772 data.}
\label{tab:alpha_psi}
\begin{tabular*}{0.3\textwidth}{@{}c@{\extracolsep{\fill}}c}
\hline
\xf\ & $\alpha^\psi$ \\ \hline
 0.15 & 0.919$\pm$0.005\\
 0.25 & 0.918$\pm$0.002\\
 0.35 & 0.910$\pm$0.004\\
 0.45 & 0.883$\pm$0.008\\
 0.55 & 0.866$\pm$0.009\\
 0.65 & 0.815$\pm$0.034\\ \hline
\end{tabular*}
\end{table}

Since the \jpsi\ cross section becomes negligible at high \xf, the
average value of $\alpha^\psi$, for positive \xf, is around
0.91.  In fact, the value $\alpha^\psi=0.920\pm0.008$ was published by
E772 in Ref.~\citen{Alde91} but the E789 collaboration uses the value
$\alpha^\psi=0.90\pm0.02$ in Ref.~\citen{Schub95}.

We move now to the p-A measurements done at the CERN/SPS by the NA38
and NA51 collaborations.  These data are certainly the best to study
the (target) nuclear effects on \jpsi\ production as a reference for
the study of the heavy ion data, since they were collected with the same
dimuon spectrometer, therefore being in the same kinematical window.
However, we should start by checking that there is compatibility
between the p-A results of NA38/51 and the observations done at
Fermilab.

The p-A results of NA38 and NA51, in what concerns \jpsi\ cross
sections, are presented in table~\ref{tab:bsig_psi_pa_exp}.  These
values correspond to the phase space domain ($\Delta$) covered by the
detector, $0<y^*<1$ and $|\cos{\theta_{CS}}|<0.5$, $y^*$ and
$\theta_{CS}$ being the center of mass rapidity and the dimuon decay
angle in the Collins-Sopper frame.

\begin{table}[hbt]
\caption{Cross sections for \jpsi\ production in p-A collisions, 
times b.r.\ into muons, as measured by the NA38 and NA51
collaborations.  See the text for details.}
\label{tab:bsig_psi_pa_exp}
\begin{tabular*}{0.7\textwidth}{@{}l@{\extracolsep{\fill}}ccc}
\hline
 & $p_\mathrm{beam}$ & $B\sigma^{\psi}_\Delta$ & $B\sigma^{\psi}_\Delta$/A \\
 & (GeV) & (nb/nucleus) & (nb/nucleon) \\ \hline
pp      & 450 & 4.41$\pm$0.52 & 4.41$\pm$0.52\\
p-$^2$H & 450 & 9.16$\pm$1.05 & 4.58$\pm$0.52\\ \hline
p-C     & 450 &46.7$\pm$5.3  & 3.89$\pm$0.44\\
p-Al    & 450 &93.8$\pm$10.7 & 3.47$\pm$0.40\\
p-Cu    & 450 &216$\pm$24   & 3.43$\pm$0.38\\
p-Cu    & 200 &104$\pm$23   & 1.65$\pm$0.37\\
p-W     & 450 &566$\pm$65   & 3.08$\pm$0.36\\
p-W     & 200 &263$\pm$18   & 1.43$\pm$0.10\\
p-U     & 200 &326$\pm$69   & 1.37$\pm$0.29\\
\hline
\end{tabular*}
\end{table}

In order to fit the \jpsi\ cross sections measured in p-A collisions
by the NA38 and NA51 collaborations to the A$^\alpha$ form, we must
first correct for the different beam energies used.  Indeed, the data
was collected either with 200~GeV or 450~GeV proton beams.  Since the
heavy ion data was collected at 200~GeV/nucleon (158~GeV in the case
of the Pb beam) we take $\sqrt{s}=19.4$~GeV as the reference and
rescale the other values.

\begin{figure}[htb]
\centering
\resizebox{0.8\textwidth}{!}{%
\includegraphics*{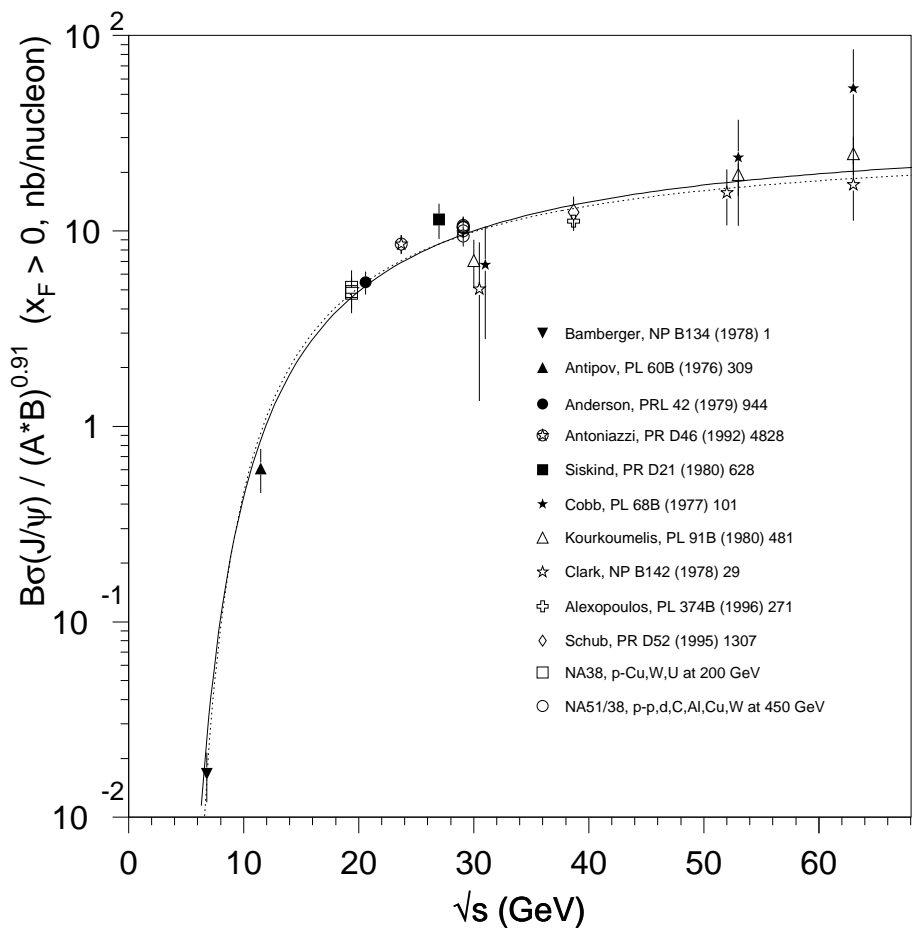}}
\caption{Energy dependence of the \jpsi\ cross section.}
\label{fig:bsig_psi_rs}
\end{figure}

The $\sqrt{s}$ dependence of the \jpsi\ cross section (times b.r.\ into
muons) is displayed in figure~\ref{fig:bsig_psi_rs}, basically taken
from fig.~11 of Ref.~\citen{Schub95} but with a few more points.  To
make this figure, the ``per nucleon'' cross sections were evaluated
using an $\alpha$ of 0.91 (it would look essentially the same with
0.92).

Notice that we have included in this figure the values measured by
NA51 and NA38 (after converting them to the positive \xf\ domain, see
below).  They are in excelent agreement among them selfs and fit very
well within the other measurements.

The solid line on the figure (the dotted line will be explained later) is 
\begin{equation}
\label{eq:rs}
{{B\sigma^{\psi}}\over{A^{0.91}}} ~(\sqrt{s}) = 37 \times
{\bigg(1-{{3.097}\over{\sqrt{s}}}\bigg)}^{12}\quad\mathrm{,~in~nb/nucleon.}
\end{equation}

We have used this function to rescale the 450~GeV values, the result
being given in table~\ref{tab:bsig_psi_pa_corr}.  In the last column
of the same table are the corresponding values in the $x_\mathrm{F}>0$
phase space domain.  We have multiplied the measured values by a
factor of 2 to account for the $\cos{\theta_{CS}}$ coverage (assuming
a uniform decay angle distribution) and by a factor 1.07 to convert
the $y*$ range into the positive \xf\ window.

\begin{table}[hbt]
\caption{Cross sections for \jpsi\ production in p-A collisions, 
times b.r.\ into muons, after corrections. See the text for
details.}
\label{tab:bsig_psi_pa_corr}
\begin{tabular*}{0.7\textwidth}{@{}l@{\extracolsep{\fill}}ccc}
\hline
 & $p_\mathrm{beam}$ & $B\sigma^{\psi}_\Delta$/A (rescaled) &
$B\sigma^{\psi}_{x_\mathrm{F}>0}$/A (rescaled)\\ 
 & (GeV) & (nb/nucleon) & (nb/nucleon) \\ \hline
pp      & 450 & 2.11$\pm$0.25 & 4.52$\pm$0.54\\
p-$^2$H & 450 & 2.19$\pm$0.25 & 4.69$\pm$0.54\\ \hline
p-C     & 450 & 1.86$\pm$0.21 & 3.98$\pm$0.45\\
p-Al    & 450 & 1.66$\pm$0.19 & 3.55$\pm$0.41\\
p-Cu    & 450 & 1.64$\pm$0.18 & 3.51$\pm$0.39\\
p-Cu    & 200 & 1.65$\pm$0.37 & 3.53$\pm$0.79\\
p-W     & 450 & 1.47$\pm$0.17 & 3.15$\pm$0.36\\
p-W     & 200 & 1.43$\pm$0.10 & 3.06$\pm$0.21\\
p-U     & 200 & 1.37$\pm$0.29 & 2.93$\pm$0.62\\
\hline
\end{tabular*}
\end{table}

From the rescaled values presented in
table~\ref{tab:bsig_psi_pa_corr}, we can fit an A$^\alpha$ function
and extract $\alpha^\psi$.  The result is 0.908$\pm$0.029, 
excluding the pp point (otherwise it is 0.919$\pm$0.021).

This is in excelent agreement with the results from Fermilab E772/E789
experiments.

\subsection{\psip\ production}

\psip\ production in p-A collisions is easier to study through the
ratio \rpsi, since the \jpsi\ has well understood energy and A
dependencies.  The values of the ratio \rpsi, ratio of production
cross-sections times branching ratios into muon pairs, measured in p-A
collisions~\cite{eps,rpsi}, are collected in table~\ref{tab:rpsi_pa}.

\begin{table}[h]
\caption{Values of the \rpsi\ ratio obtained in p-A collisions.}
\label{tab:rpsi_pa}
\begin{tabular*}{0.7\textwidth}{@{}l@{\extracolsep{\fill}}ccc}
\hline
 & $p_\mathrm{beam}$~(GeV) &
${B\sigma}^{\psi'}\,/\,{B\sigma}^{\psi}$~(\%)& Exp.t \\ 
\hline
p--H$_2$ & 450 &$ 1.69\pm 0.03 $& NA51 \\
p--D$_2$ & 450 &$ 1.80\pm 0.03 $& NA51 \\
p--C     & 450 &$ 1.90\pm 0.13 $& NA38 \\
p--Al    & 450 &$ 1.36\pm 0.35 $& NA38 \\
p--Cu    & 450 &$ 1.68\pm 0.11 $& NA38 \\
p--W     & 450 &$ 1.59\pm 0.13 $& NA38 \\
p--W     & 200 &$ 1.80\pm 0.17 $& NA38 \\
p--U     & 200 &$ 1.77\pm 0.22 $& NA38 \\
\hline
p--p  & $\sqrt{s}=63$ &$ 1.9\pm 0.6 $&ISR\\
p--Li & 300 &$ 1.88\pm 0.26\pm0.05 $&E\,705\\
p--Be & 400 &$ 1.7\pm 0.5 $&E\,288\\
p--Si & 800 &$ 1.65\pm 0.20 $&E\,771\\ 
p--Au & 800 &$ 1.80\pm 0.1\pm0.2 $&E\,789\\ 
\hline
\end{tabular*}
\end{table}

These values show that the ratio between \psip\ and \jpsi\ production
cross sections is independent of the target nuclei and (at high
energies) of $\sqrt{s}$.  In fact, we can even add values obtained
with other beam particles~\cite{eps,E687} ($\pi^+$, $\pi^-$,
$\bar{\mathrm{p}}$, $\gamma$) without changing the picture.

This observation is a clear indication that these charmonia bound
states are actually only formed as such outside the target nucleus, at
least when they are fast enough (at positive $x_F$).  It would
certainly be very interesting to see what happens when the $c\bar{c}$
state is slow enough to become a \jpsi\ or \psip\ within nuclear
matter~\cite{invkin}.  Present (p-A) dimuon experiments are unable to
detect the correspondingly slow decay muons but this might soon be
overcome by an ``inverse kinematics experiment'', sending the SPS Pb
beam on a light target.
\begin{figure}[htp]
\centering
\resizebox{0.73\textwidth}{!}{%
\includegraphics*{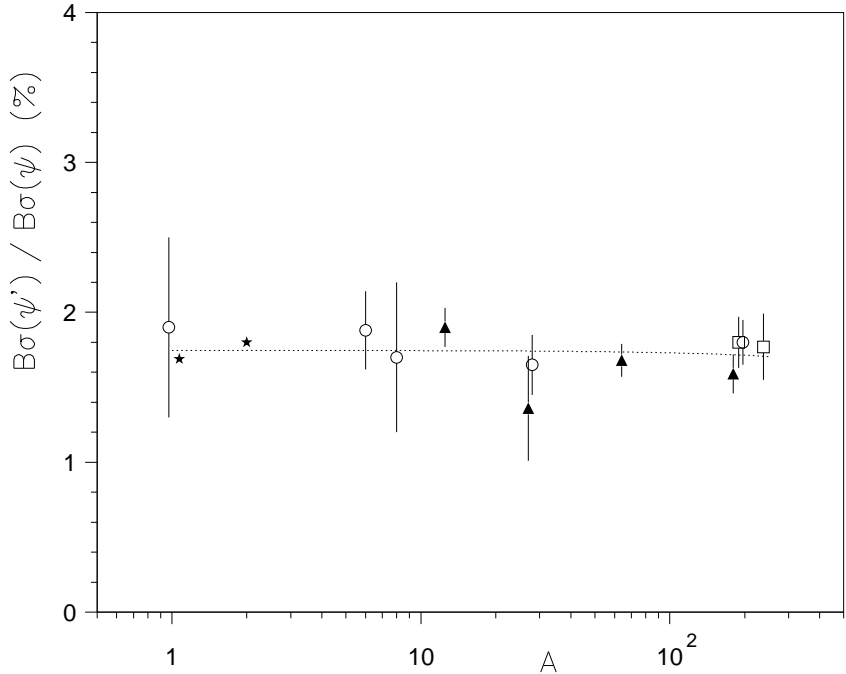}}
\caption{Target dependence of the \rpsi\ ratio, in p-A collisions.}
\label{fig:rpsi_a}
\end{figure}

\begin{figure}[htb]
\centering
\vspace{-0.5cm}
\resizebox{0.73\textwidth}{!}{%
\includegraphics*{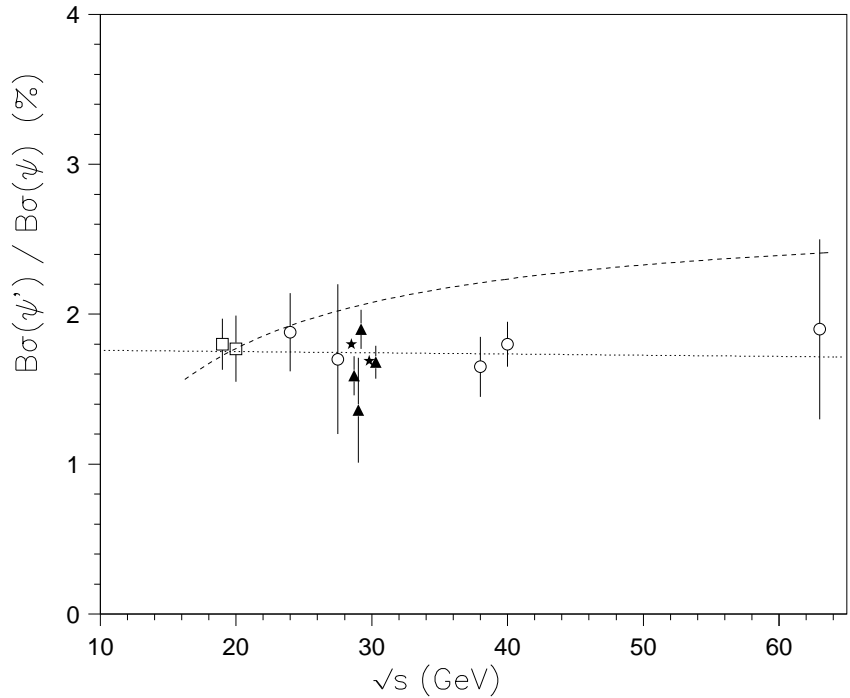}}
\caption{Energy dependence of the \rpsi\ ratio, in p-A collisions.}
\label{fig:rpsi_rs}
\end{figure}

The data of table~\ref{tab:rpsi_pa} have been plotted versus A in
figure~\ref{fig:rpsi_a} and versus \rs\ in figure~\ref{fig:rpsi_rs}.
In these figures, some of the points were slightly displaced in the
horizontal axis (otherwise a few points would be indistinguishable).

The dotted curves in these plots are the result from linear fits to
the data, the slopes being compatible with zero for both the A and the
\rs\ cases, $(-1.6\pm4.2)\times10^{-4}$ and
$(-8.1\pm79)\times10^{-4}$, respectively.

Even if $\alpha^{\psi'}$ would be smaller than $\alpha^{\psi}$ by only
0.01, that would imply a decrease of \rpsi\ from pp to p-U of 5.6\,\%,
already quite difficult to accomodate by these measurements.
Therefore, models which predict a substantially different A dependence
for the \jpsi\ and \psip\ resonances, in p-A collisions, are in clear
contradiction with the data.

Figure~\ref{fig:rpsi_rs} also contains a second (dashed) curve, which
represents the evolution with \rs\ of the \rpsi\ ratio expected if we
assume the \psip\ cross section to depend on \rs\ according to
equation~\ref{eq:rs}, with the parameter 3.097~GeV replaced by
3.685~GeV.  The fact that this curve (normalised in the plot at
$\sqrt{s}=19.4$~GeV) completely fails to go through the data points,
shows very well how incorrect such an assumption is, and reveals that
this parameter should not be directly associated to the mass of the
particle under study.  In fact, the best fit to the points in
figure~\ref{fig:bsig_psi_rs} is provided by
$31.5\times{(1-4.4/\sqrt{s})}^{7.3}$ (dotted line in that figure).

\section{Hard processes in nucleus-nucleus collisions}

\dy\ differential cross sections (corrected for detector specific
acceptance and smearing effects) are not yet available for
nucleus-nucleus collisions.  Work in this direction is still in progress
within the NA38/50 collaborations.  Therefore, it is not yet possible
to compare the data with NLO calculations (not available in the
form of event generators).  Although the nuclear
effects on the parton distribution functions are expected to be small,
in the kinematical window of NA38/50, they are certainly non-zero and
it would be very important to actually measure them rather than just
neglect their existence.  For the moment, the fact that the same K
factor (relative to a certain LO calculation) is required by the p-A
and ion data~\cite{Gonin} is the best argument to say that \dy\ is a
good reference in the studies of charmonia production by ion
collisions.

The \jpsi\ cross sections measured in nucleus-nucleus collisions, by
the NA38 and NA50 collaborations, are collected in 
table~\ref{tab:bsig_psi_ba}, before and after \rs\ and phase space
corrections.  

\begin{table}[ht]
\caption{\jpsi\ cross sections, times b.r.\ into muons,
  measured by the NA38 and NA50 collaborations in B-A collisions.}
\label{tab:bsig_psi_ba}
\begin{tabular*}{0.95\textwidth}{@{}l@{\extracolsep{\fill}}cccc}
\hline
 & $p_\mathrm{beam}$ 
& $B\sigma^{\psi}_\Delta$ 
& $B\sigma^{\psi}_\Delta$/(B$\times$A) (rescaled)
& $B\sigma^{\psi}_{x_\mathrm{F}>0}$/(B$\times$A) (rescaled)\\ 
 & (GeV) & ($\mu$b) & (nb/nucleon) & (nb/nucleon) \\ \hline
O-Cu & 200 &1.26$\pm$0.13 &1.25$\pm$0.13 & 2.68$\pm$0.28 \\
O-U  & 200 &4.42$\pm$0.46 &1.16$\pm$0.12 & 2.48$\pm$0.26 \\
S-U  & 200 &7.69$\pm$0.76 &1.01$\pm$0.10 & 2.16$\pm$0.21 \\ \hline
Pb-Pb& 158 &19.0$\pm$1.4  &0.59$\pm$0.04 & 1.26$\pm$0.09 \\ \hline
\end{tabular*}
\end{table}

\begin{figure}[!h]
\centering
\vspace{-0.3cm}
\resizebox{0.6\textwidth}{!}{%
\includegraphics*{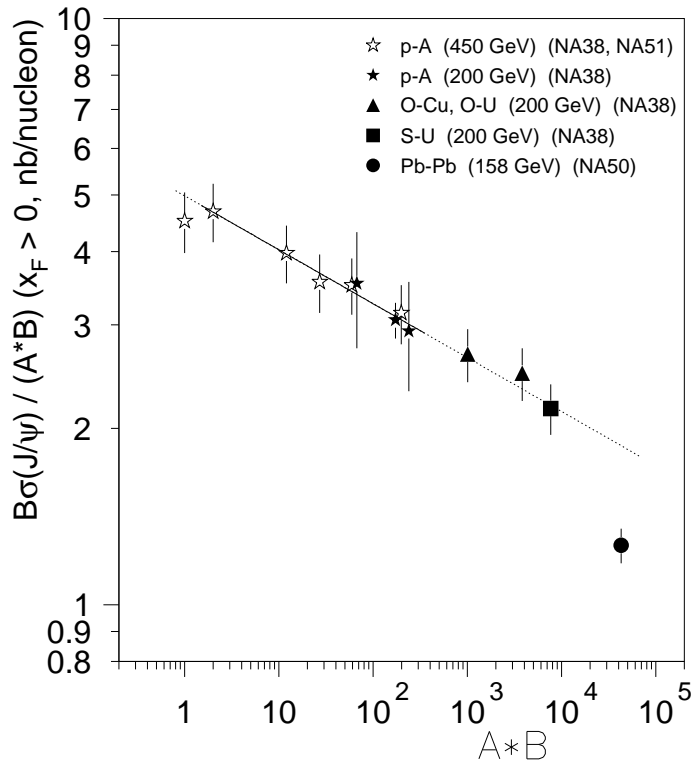}}
\vspace{-0.3cm}
\caption{\jpsi\ production cross sections (times b.r.) versus
  A$\times$B.  The straight line, in this log-log plot, corresponds to
  $\alpha=0.908$, the dotted part being an extrapolation from the
  fitted region.}
\label{fig:bsig_psi_ab}
\end{figure}

The light ion collisions are in excelent agreement with the
extrapolation of the proton data ``nuclear absorption'' curve.
Indeed, if we include the oxygen and sulphur points in the fit to a
(B$\times$A)$^\alpha$ function, the value of $\alpha$ stays the same
(0.911$\pm$0.016) as before.  This smooth transition from p-A to
O,\,S-A collisions is illustrated in figure~\ref{fig:bsig_psi_ab}.

Contrary to what happens in the light ion region, the Pb-Pb point is
seen to be away from the ``expected'' production cross section, which
is 50\,\% higher than the measured value.  This is a strong indication
that something else besides ``nuclear absorption'' is happening to the
\jpsi\ in Pb-Pb collisions.  

There are more precise ways of checking whether nuclear absorption
alone can explain the observed evolution of \jpsi\ production yields.
On the vertical axis we can use the ratio of \jpsi\ to \dy\ yields to
reduce the systematic errors resulting from absolute normalisations.
To explore the dependence of \jpsi\ production on the centrality of
the ion collisions, we can subdivide the event samples according to
some global variable, like \et\ or \ezdc, sensitive to the geometry of
the collision.  In fact, the best variable within the ``nuclear
absorption'' framework is the amount of nuclear matter crossed by the
pre-\jpsi\ state on its way out, the famous L introduced by Gerschel
and H\"ufner~\cite{Gerschel_Huefner}.

\begin{figure}[hb]
\centering
\resizebox{0.7\textwidth}{!}{%
\includegraphics*{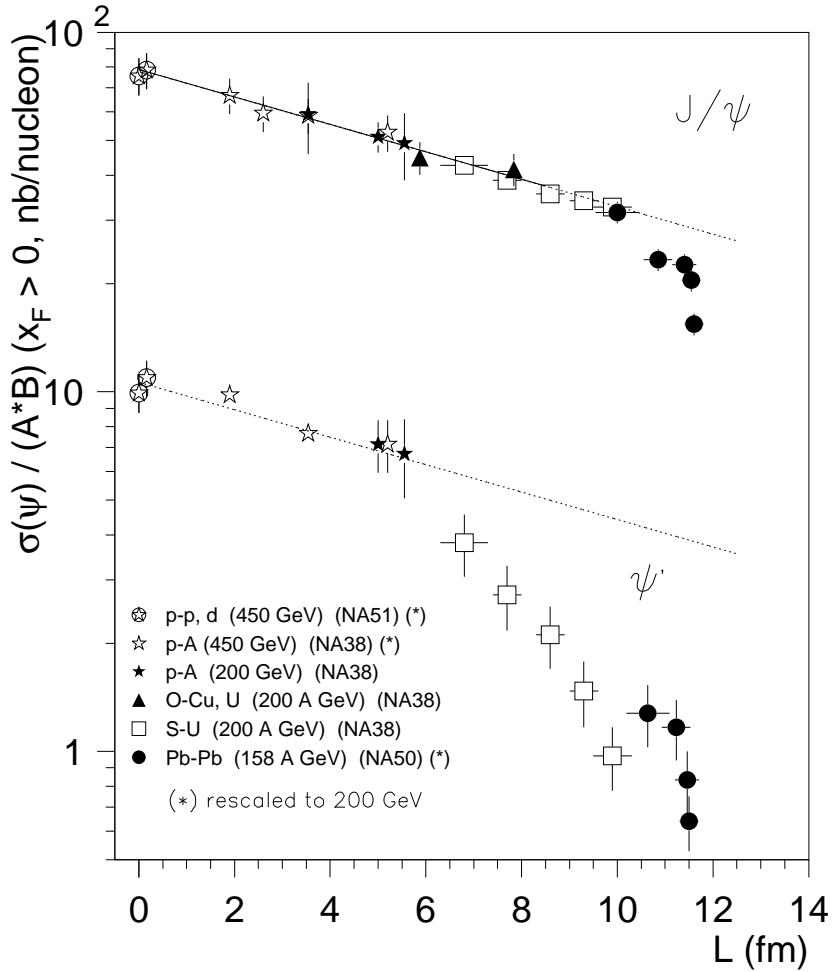}}
\caption{\jpsi\ and \psip\ production cross sections versus L.
  The (parallel) lines correspond to $\exp(-\rho\sigma \mathrm{L})$, with
  $\rho\sigma=0.088$~fm$^{-1}$.}
\label{fig:sigpsi_l}
\end{figure}

Figure~\ref{fig:sigpsi_l} attempts to merge in a single plot the
variation with L 
of both the \jpsi\ and \psip\ cross sections.  To
cope with the range in the vertical scale, the branching ratios into
muons were corrected for.  The (S-U and Pb-Pb) L dependent
$\sigma^\psi$ values were obtained from \psidy\ using the all-\et\ 
values for normalisation and assuming that DY does not depend on L (in
other words, the integrated point is replaced by 5 points keeping the
same average).  The $\sigma^{\psi'}$ points were obtained from the
\rpsi\ values (with some art to combine the different centrality bins
used in the \jpsi\ and \psip\ analyses~\cite{Gonin}).

We can see that the nuclear absorption model reproduces quite well all
the \jpsi\ points up to the most peripheral Pb-Pb collisions.  The
remaining Pb-Pb points reveal the existence of a new ``suppression''
mechanism, L not being any longer an appropriate parameter.

The \psip\ points reveal that after L~$\sim$~6~fm the two charmonia
states are already fully formed and start feeling in a different way
the medium they cross.  The nuclear matter path length alone is no
longer enough to correctly parametrise the \psip\ behaviour, revealing
that a dense hadronic medium is formed in S-U collisions.

To understand whether the new Pb-Pb data on charmonia production is
more than just a ``hint'' of QGP formation, accurate studies will have
to be pursued, much beyond the ``L framework''.

For instance, a correct understanding of the S-U data must consider
that about 8\,\% of the observed \jpsi\ yield actually come from
\psip\ decays.  Since the \psip\ is significantly suppressed in S-U
collisions, a certain fraction of the \jpsi\ yield must also be
suppressed.

To conclude, table~\ref{tab:psi_et} collects useful information for
further studies of \jpsi\ production from p-A to S-U and Pb-Pb
collisions, as a function of \et.  Notice, however, that the \et\ 
scales are not the same in the NA38 and NA50 experiments, due to the
different rapidity windows covered by the calorimeters: $1.7<\eta<4.1$
in NA38 and $1.1<\eta<2.3$ in NA50.  The resolution of this
non-trivial problem would require a few more pages.

\begin{table}[hbt]
\caption{\et\ dependence of \jpsi\ yields:  \psidy\ and after
  normalisation to the integrated cross section and to the pp value.}
\label{tab:psi_et}
\begin{tabular*}{0.85\textwidth}{@{}lc@{\extracolsep{\fill}}ccc}
\hline
 & \et & \psidy & $B\sigma^{\psi}_{\mathrm{BA}}$\,/\,(B\,A)
& $\sigma^{\psi}_{\mathrm{BA}}$\,/\,(B\,A\,$\sigma^{\psi}_{\mathrm{pp}}$)\\
 & (GeV) & & (nb/nucleon) & \\ 
\hline
S-U & All & 21.4$\pm$0.24 & 1.01$\pm$0.10 & 0.48$\pm$0.07 \\
\hline
 & 25 & 25.2$\pm$0.69& 1.19$\pm$0.12& 0.56$\pm$0.09\\
 & 42 & 22.8$\pm$0.56& 1.08$\pm$0.11& 0.51$\pm$0.08\\
 & 57 & 21.0$\pm$0.48& 0.99$\pm$0.10& 0.47$\pm$0.07\\
 & 71 & 20.2$\pm$0.44& 0.95$\pm$0.10& 0.45$\pm$0.07\\
 & 82 & 19.2$\pm$0.44& 0.91$\pm$0.09& 0.43$\pm$0.07\\
\hline
Pb-Pb & All & 11.9$\pm$0.4 & 0.59$\pm$0.04 & 0.28$\pm$0.04 \\
\hline
 & 35 &17.8$\pm$2.20 &0.88$\pm$0.13 &0.42$\pm$0.08\\
 & 59 &13.2$\pm$0.99 &0.65$\pm$0.07 &0.31$\pm$0.05\\
 & 88 &12.7$\pm$0.82 &0.63$\pm$0.06 &0.30$\pm$0.05\\
 &120 &11.4$\pm$0.80 &0.57$\pm$0.06 &0.27$\pm$0.04\\
 &149 & 8.6$\pm$0.80 &0.43$\pm$0.05 &0.20$\pm$0.03\\
\hline
\end{tabular*}
\end{table}

\section*{Acknowledgements}

It is a pleasure to acknowledge fruitful discussions with
J.-P.~Blaizot, K.J.~Eskola, D.~Kharzeev, P.L.~McGaughey, A.~Morsch,
P.V.~Ruuskanen, H.~Satz, J.~Schukraft, G.~Schu\-ler, K.~Sridhar,
C.Y.~Wong and with my colleagues from the NA38, NA51 and NA50
collaborations.  Very special thanks are due to Dima, Helmut and Pat.


\begin{thebibliography}{99}

\bibitem{Matsui_Satz} T.~Matsui and H.~Satz, Phys. Lett. \textbf{B178}
  (1986) 416.

\bibitem{Bussiere87} A.~Bussiere \emph{et al.} (NA38 Coll.),
  Z.~Phys.~\textbf{C38} (1988) 117 (QM~'87).

\bibitem{Baglin91a} C.~Baglin \emph{et al.} (NA38 Coll.),
  Phys. Lett. \textbf{B255} (1991) 459.

\bibitem{NA3_psi} J.~Badier \emph{et al.} (NA3 Coll.),
  Z.~Phys.~\textbf{C20} (1983) 101. 
  
\bibitem{NA38_et} C.~Baglin \emph{et al.} (NA38 Coll.),
  Phys. Lett. \textbf{B251} (1990) 472.
  
\bibitem{Sridhar96} See, for example, the contribution of K.~Sridhar
  to the ``31st Rencontres de Moriond: QCD and High Energy Hadronic
  Interactions'', March 1996 (hep-ph/9606252).

\bibitem{CDF} A.~Sansoni \emph{et al.} (CDF Coll.), these proceedings
  (FERMILAB-CONF-96-221-E).
  
\bibitem{Braaten} E.~Braaten, these proceedings
  (hep-ph/9608370).\\
  E.~Braaten, S.~Fleming and T.C.~Yuan, OHSTPY-HEP-T-96001
  (hep-ph/9602374).

\bibitem{Kharzeev_Satz96} D.~Kharzeev and H.~Satz,
  Phys. Lett. \textbf{B366} (1996) 316 (hep-ph/9508276).

\bibitem{Gerschel_Huefner} C.~Gerschel and J.~H\"ufner,
  Z.~Phys.~\textbf{C56} (1992) 171.

\bibitem{Kharzeev_Satz94} D.~Kharzeev and H.~Satz,
  Phys. Lett. \textbf{B334} (1994) 155.
  
\bibitem{Gonin} M.~Gonin \emph{et al.} (NA50 Coll.), these proceedings.
  
\bibitem{IMR} C.~Louren\c{c}o \emph{et al.} (NA38 Coll.), Nucl. Phys.
  \textbf{A566} (1994) 77c (QM~'93).\\
  E.~Scomparin \emph{et al.} (NA50 Coll.), these proceedings.
  
\bibitem{E772_dy} P.L.~McGaughey \emph{et al.} (E772 Coll.), Phys.
  Rev. \textbf{D50} (1994) 3038.

\bibitem{E605_dy} G.~Moreno \emph{et al.} (E605 Coll.),
  Phys. Rev. \textbf{D43} (1991) 2815. 
  
\bibitem{NA3_dy} J.~Badier \emph{et al.} (NA3 Coll.),
  Z.~Phys.~\textbf{C26} (1984) 489. 

\bibitem{hepdata} http://durpdg.dur.ac.uk/scripts/help.csh/REAC/EXP/LISTREACEXP

\bibitem{Pat} P.L.~McGaughey, these proceedings and references
  therein.
  
\bibitem{dy_th} W.L.~van Neerven, Int. J. Mod.  Phys. \textbf{A10}
  (1995) 2921.\\
  S.~Gavin \emph{et al.} (Hard Probe Coll.), Int. J. Mod.  Phys.
  \textbf{A10} (1995) 2961.
  
\bibitem{udbar} A.~Baldit \emph{et al.} (NA51 Coll.), Phys. Lett.
  \textbf{B332} (1994) 244.

\bibitem{NMC} P.~Amaudruz \emph{et al.} (NMC Coll.), Phys. Rev. Lett.
  \textbf{66} (1991) 2712.

\bibitem{Alde91} D.M.~Alde \emph{et al.} (E772 Coll.), Phys. Rev. Lett.
  \textbf{66} (1991) 133.

\bibitem{Schub95} M.H.~Schub \emph{et al.} (E789 Coll.),
  Phys. Rev. \textbf{D52} (1995) 1307.
  
\bibitem{eps} C.~Louren\c{c}o \emph{et al.} (NA38/50 Coll.), Proc. of
  the Int. Europhysics Conf. on HEP, Brussels 1995, p.\,363
  (CERN-PRE-95-001).

\bibitem{rpsi}
  H.D.~Snyder \emph{et al.} (E288 Coll.),
  Phys. Rev. Lett. \textbf{36} (1976) 1415.\\
  A.G.~Clark \emph{et al.},
  Nucl. Phys. \textbf{B142} (1978) 29.\\
  L.~Antoniazzi \emph{et al.} (E705 Coll.), Phys. Rev. \textbf{D46}
  (1992) 4828.\\
  M.H.~Schub \emph{et al.} (E789 Coll.), Phys. Rev. \textbf{D52}
  (1995) 1307.\\
  T.~Alexopoulos \emph{et al.} (E771 Coll.), Phys. Lett. \textbf{B374}
  (1996) 271.
  
\bibitem{E687} P.L.~Frabetti \emph{et al.} (E687 Coll.), ``\jpsi\ and
  \psip\ Photoproduction in E687'', comm. at the Int.  Europhysics
  Conf. on HEP, Brussels 1995, (EPS0705), and ref.s therein.
  
\bibitem{invkin} D.~Kharzeev and H.~Satz, Phys. Lett. \textbf{B356}
  (1995) 365.

\end{thebibliography}
\end{document}